\documentclass[aps,graphicx,twocolumn]{revtex4}%showkeys
\usepackage{amssymb}
\usepackage{amsmath}
\usepackage{graphicx}
\usepackage{array}
\usepackage{color}
\usepackage{epstopdf}

\begin{document}

\title{General quantum computation on photons assisted with double single-sided cavity system}

\author{Jiu-Ming Li$^{1}$}\email{ljm_physics2020@126.com}
\author{Jun-Yan Liu$^{1}$}
\author{Yuan-Yuan Liu$^{1}$}
\author{Xiao-Ming Xiu$^{1}$}
\author{Shao-Ming Fei$^{2}$}\email{feishm@cnu.edu.cn}

\address{$^{1}$College of Physics Science and Technology, Bohai University, Jinzhou 121013, China   \\
$^{2}$School of Mathematical Sciences, Capital Normal University, Beijing 100048, China}

\begin{abstract}
We propose a physical system consisting of two optical cavities and a two-level system (TLS), which can be viewed as a double single-sided cavity system. The two cavities are crossed each other in a mutually perpendicular way and are both single-sided. The TLS is coupled to the two cavities. The universal input-output relation of the system, the reflection and transmission coefficients are derived by exploiting the probability amplitude method. Then by using the nitrogen-vacancy center instead of the TLS, we generate the controlled-phase gate and the controlled-controlled-phase gate on the photon qubits, with simple protocols that can be accomplished in both weak and strong coupling regimes. The protocols are shown to give rise to high fidelities and gate efficiencies.
\end{abstract}

%\keywords{}

\maketitle

\section{INTRODUCTION}%(I)

By exploiting the principles of quantum mechanics, quantum computation \cite{QC1,QC2} can be much faster and more efficient than classical computation. The key issues in quantum computation are the realizations of quantum logic gates in quantum physical systems. Among these quantum logic gates, the controlled-phase gate \cite{cp1,cp2,cp3,cp4} and controlled-controlled-phase gate \cite{ccp1,ccp2} are the important two-qubit and three-qubit quantum gates. Currently, many quantum systems have been proposed for accomplishing certain quantum computation, such as cavity quantum electrodynamics (QED) \cite{cQED1,cQED2,cQED3}, circuit-QED \cite{cir1,cir2,cir3}, waveguide-QED \cite{wave1,wave2}, cavity magnonics \cite{cm1,cm2,cm3,cm4,cm5,cm6}, quantum dots \cite{qd1,qd2,qd3,qd4}, trapped ions \cite{ti1,ti2,ti3}, nitrogen-vacancy centers \cite{NV1,NV2,NV3,NV4,NV5,NV6,NV7} and Rydberg atoms \cite{Ra1,Ra2,Ra3,Ra4}.

In recent years, many quantum information tasks are accomplished by exploiting the input-output relations of specific systems combined with other systems \cite{IO1,IO2,IO3,IO4,IO5,IO6,IO7,IO8,IO9}, like a cavity coupled to a bath. The cavity is a single-mode field, and the bath fields are viewed as a series of harmonic oscillator states. Through the coupling mirror of the cavity, the input-output relation of the cavity coupled to the bath is attained by exploiting the Heisenberg equations of motion (HEM) for the field operators. In some protocols of quantum computation, there are mainly two kinds of systems. One is the single-sided cavity system \cite{IO2,IO4,IO5,IO6,IO8,IO9}, where the cavity is coupled to the bath through only one coupling mirror. So the single-sided cavity has only a reflection coefficient. Another is the double-sided cavity system \cite{IO1,IO3,IO7}. The cavity of the system is coupled to the bath through both coupling mirrors, and the system has a reflection coefficient and a transmission coefficient at the same time. However, the similar input-output relation can be obtained by exploiting probability amplitude method (PAM) \cite{PAM}. In the system of a cavity coupled to a bath, assumed that a single photon is incident from the ports of the cavity, a set of kets represented the evolved system can be obtained. Through the PAM, the relation between the field amplitudes is calculated. Besides, the reflection coefficient for single-sided cavity or the reflection and transmission coefficients for the double-sided cavity can be attained. Compared with HEM, the picture given by the PAM is not changed in the whole process.

In this work, we first consider a system consisting of two optical cavities and a two-level system (TLS). The two optical cavities are crossed each other in a mutually perpendicular way, and the TLS is coupled to the two cavities simultaneously. In this system, the two cavities are both single-sided, namely, each cavity is coupled to the corresponding port through the coupling mirror. So the system can be viewed as a double single-sided cavity system. By exploiting the PAM in the single-excitation subspace, the universal input-output relation of the system is obtained, and the reflection and transmission coefficients of the system are attained and analyzed. Then, we use the nitrogen-vacancy (NV) center instead of the TLS in the system. The relations between the states of the incident photons and the states of the NV-center is established. At last, we construct the controlled-phase (c-phase) gate and the controlled-controlled-phase (cc-phase) gate on the incident photons. In the protocol of the c-phase gate, the different circularly polarized lights of two incident photons are used through the polarizing beam splitter (PBS), respectively. For the generation of the cc-phase gate, the states of the two incident photons and the two NV-centers in corresponding systems are prepared firstly, in which the two photons are viewed as the control qubits. Then, a third photon is incident and interacted with the whole system. The cc-phase gate is accomplished on three incident photons at last. Compared with other works, we obtain the universal input-output relation of the double single-sided cavity system, the reflection and transmission coefficients, as well as the c-phase gate and the cc-phase gate with much simpler protocols.

This work is organized as follows. In Sec. \ref{S2}, we present the universal input-output relation of the system. The protocols of the c-phase gate and the cc-phase gate are given in Sec. \ref{S3}. In Sec. \ref{S4}, we analyze the fidelities and the efficiencies of the protocols. We conclude in Sec. \ref{S5}.

\section{THE UNIVERSAL INPUT-OUTPUT RELATION OF THE DOUBLE SINGLE-SIDED CAVITY SYSTEM}\label{S2}%(II)

\begin{figure}[!htbp]
\centering
\includegraphics[width=8.0cm,angle=0]{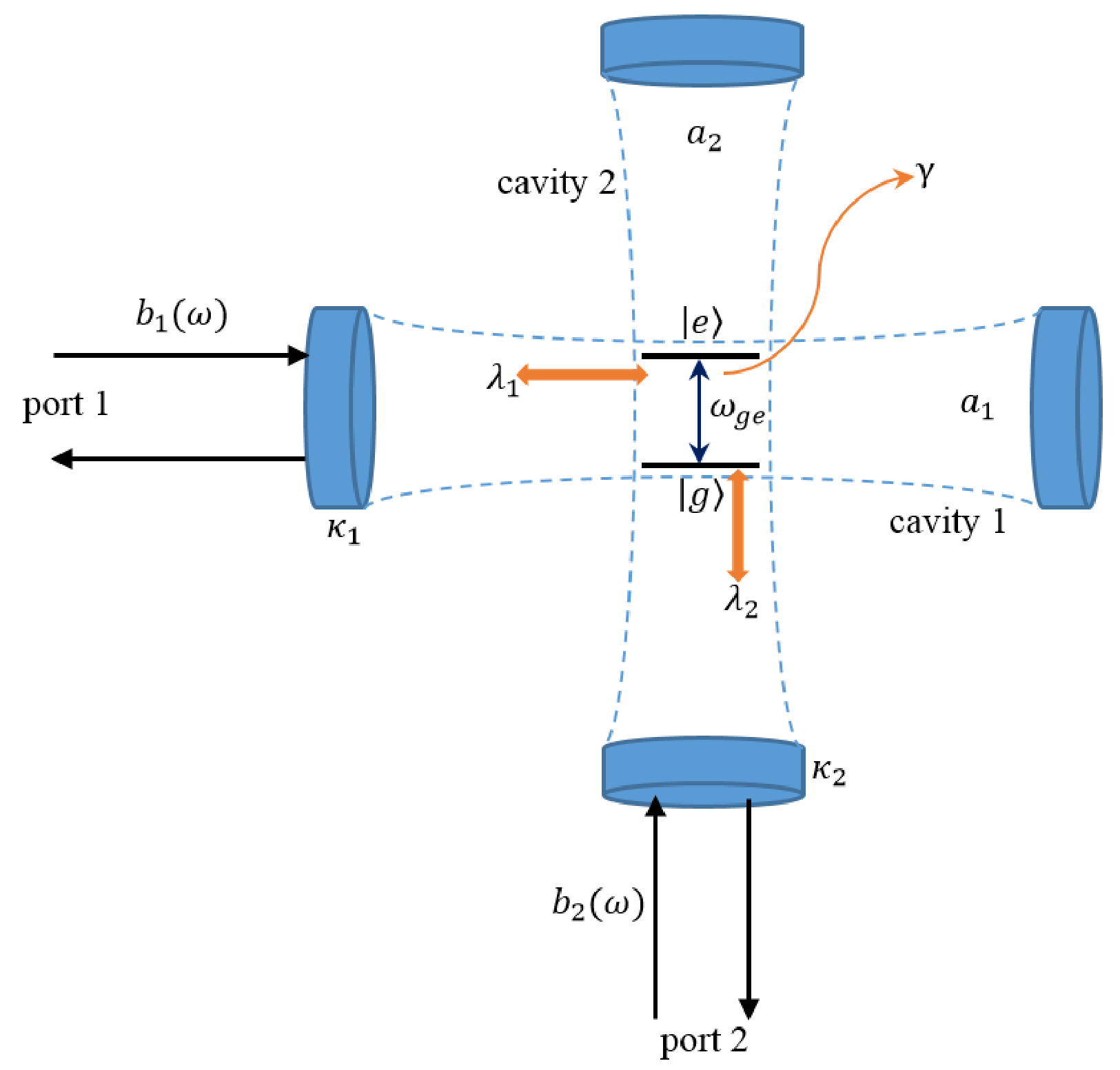}
\caption{(Color online) Schematic of the double single-sided cavity system composed of two optical cavities and a TLS. The two cavities are both single-sided and coupled to the corresponding ports through the coupling mirrors. The TLS is interacted with the two cavities simultaneously.}\label{fig1}
\end{figure}

Consider a system consisting of two optical cavities and a TLS, as shown in Fig. \ref{fig1}, which is viewed as a double single-sided cavity system, with the two cavities being crossed each other in a mutually perpendicular way, and the TLS is coupled to the two cavities simultaneously. The two cavities, labeled as cavity 1 and cavity 2, are both single-sided and coupled to the port 1 and 2 through the coupling mirror, respectively.

Let $|g\rangle$ and $|e\rangle$ be the ground and excited states of the TLS, respectively. In the Schr\"{o}dinger picture, the Hamiltonian of the system is written as (hereafter $\hbar=1$),
\begin{eqnarray}
H\!\!&=&\!\!(\omega_{ge}-\mathrm{i}\frac{\gamma}{2})|e\rangle\langle e|+\omega_1a_1^\dag a_1+\omega_2a_2^\dag a_2   \nonumber    \\
&&\!\!+\int_{-\infty}^{+\infty}\!\!\!\mathrm{d}\omega\,\omega b_1^\dag(\omega)b_1(\omega)+\int_{-\infty}^{+\infty}\!\!\!\mathrm{d}\omega\,\omega b_2^\dag(\omega)b_2(\omega)        \nonumber    \\
&&\!\!+\mathrm{i}\sqrt{\frac{\kappa_1}{2\pi}}\int_{-\infty}^{+\infty}\mathrm{d}\omega\big[a_1b_1^\dag(\omega)-a_1^\dag b_1(\omega)\big]      \nonumber     \\
&&\!\!+\mathrm{i}\sqrt{\frac{\kappa_2}{2\pi}}\int_{-\infty}^{+\infty}\mathrm{d}\omega\big[a_2b_2^\dag(\omega)-a_2^\dag b_2(\omega)\big]      \nonumber     \\
&&\!\!+\mathrm{i}\lambda_1(a_1\sigma^+-a_1^\dag \sigma)+\mathrm{i}\lambda_2(a_2\sigma^+-a_2^\dag \sigma).
\end{eqnarray}
Here, $a_1$ and $a_1^\dag$ ($a_2$ and $a_2^\dag$) are the annihilation and creation operators of the cavity 1 (cavity 2) satisfying $[a_1, a_1^\dag]=1$ ($[a_2, a_2^\dag]=1$). $b_1(\omega)$ ($b_2(\omega)$) and $b_1^\dag(\omega)$ ($b_2^\dag(\omega)$) denote the annihilation and creation operators of the port 1 (port 2) satisfying $[b_1(\omega), b_1^\dag(\omega^\prime)]=\delta(\omega-\omega^\prime)$ ($[b_2(\omega), b_2^\dag(\omega^\prime)]=\delta(\omega-\omega^\prime)$). $\sigma=|g\rangle\langle e|$ and $\sigma^+=|e\rangle\langle g|$ are the lowing and raising operators of the TLS. $\omega_1$ and $\omega_2$ are the oscillating frequencies of the two optical cavities. $\omega_{ge}$ is the transition frequency of the TLS. $\sqrt{\kappa_1/2\pi}$ ($\sqrt{\kappa_2/2\pi}$) is the coupling strength between the cavity 1 (2) and the port 1 (2), and $\kappa_1$ ($\kappa_2$) is the decay rate of the coupling mirror of the cavity 1 (2). $\gamma$ is the spontaneous emission rate of the TLS. $\lambda_1$ ($\lambda_2$) is the coupling strength between the TLS and the cavity 1 (2).

In the single-excitation subspace, we have five kets, $|1,\emptyset,0,0,g\rangle$, $|\emptyset,1,0,0,g\rangle$, $|\emptyset,\emptyset,1,0,g\rangle$, $|\emptyset,\emptyset,0,1,g\rangle$ and $|\emptyset,\emptyset,0,0,e\rangle$, labeled as $|\varepsilon\rangle_1$, $|\varepsilon\rangle_2$, $|\varepsilon\rangle_3$, $|\varepsilon\rangle_4$ and $|\varepsilon\rangle_5$, respectively. Each ket denotes the number states of the port 1 and the port 2, the Fock states of the cavity 1 and the cavity 2, and the state of the TLS. Thus, if a single photon is incident from port 1 or port 2, the state of the system evolves to
\begin{eqnarray}
|\psi(t)\rangle\!\!&=&\!\!\int\mathrm{d}\omega\,A(\omega,t)|\varepsilon\rangle_1
+\int\mathrm{d}\omega\,B(\omega,t)|\varepsilon\rangle_2                 \nonumber     \\
&&\!\!+C(t)|\varepsilon\rangle_3+D(t)|\varepsilon\rangle_4+E(t)|\varepsilon\rangle_5
\end{eqnarray}
at time $t$, where $A(\omega,t)$, $B(\omega,t)$, $C(t)$, $D(t)$ and $E(t)$ are the probability amplitudes of the corresponding kets.

According to the Schr\"{o}dinger equation, five differential equations are attained as followed
\begin{eqnarray}
&&\dot{A}(\omega,t)=-\mathrm{i}\omega A(\omega,t)+\sqrt{\frac{\kappa_1}{2\pi}}C(t),      \nonumber    \\
&&\dot{B}(\omega,t)=-\mathrm{i}\omega B(\omega,t)+\sqrt{\frac{\kappa_2}{2\pi}}D(t),      \nonumber    \\
&&\dot{C}(t)=-\mathrm{i}\omega_1C(t)-\sqrt{\frac{\kappa_1}{2\pi}}\int\mathrm{d}\omega\,A(\omega,t)-\lambda_1E(t),         \nonumber  \\
&&\dot{D}(t)=-\mathrm{i}\omega_2D(t)-\sqrt{\frac{\kappa_2}{2\pi}}\int\mathrm{d}\omega\,B(\omega,t)-\lambda_2E(t),          \nonumber   \\
&&\dot{E}(t)=(-\mathrm{i}\omega_{ge}-\frac{\gamma}{2})E(t)+\lambda_1C(t)+\lambda_2D(t).      \label{eq1}
\end{eqnarray}
The input amplitudes of the two ports are defined by
\begin{eqnarray}
&&A_{\mathrm{in}}(t)=\frac{1}{\sqrt{2\pi}}\int\mathrm{d}\omega\,A(\omega,0)\mathrm{e}^{-\mathrm{i}\omega t},   \nonumber      \\
&&B_{\mathrm{in}}(t)=\frac{1}{\sqrt{2\pi}}\int\mathrm{d}\omega\,B(\omega,0)\mathrm{e}^{-\mathrm{i}\omega t}.
\end{eqnarray}
By Fourier transformation we obtain
\begin{eqnarray}
&&C(\omega)=\frac{-\sqrt{\kappa_1}A_{\mathrm{in}}(\omega)-\lambda_1E(\omega)}{\frac{1}{2}\kappa_1-\mathrm{i}\delta_1},  \nonumber    \\
&&D(\omega)=\frac{-\sqrt{\kappa_2}B_{\mathrm{in}}(\omega)-\lambda_2E(\omega)}{\frac{1}{2}\kappa_2-\mathrm{i}\delta_2},  \nonumber    \\
&&E(\omega)=\frac{\lambda_1C(\omega)+\lambda_2D(\omega)}{\frac{1}{2}\gamma-\mathrm{i}\delta_{ge}},
\end{eqnarray}
where $\delta_1=\omega-\omega_1$, $\delta_2=\omega-\omega_2$ and $\delta_{ge}=\omega-\omega_{ge}$.

We first consider that the TLS is coupled to the two cavities simultaneously, namely, $\lambda_1\neq0$ and $\lambda_2\neq0$. Assumed that a single photon is incident from the port 1. Given the input-output relation of the cavity 1 and port 1, $A_{\mathrm{out}}(\omega)=A_{\mathrm{in}}(\omega)+\sqrt{\kappa_1}C(\omega)$, the universal input-output relation of the double single-sided cavity system is given by $A_{\mathrm{out}}(\omega)=r(\omega)A_{\mathrm{in}}(\omega)+t(\omega)B_{\mathrm{in}}(\omega)$, where
\begin{widetext}
\begin{eqnarray}
&&r(\omega)=\frac{\left[\left(-\frac{\kappa_1}{2}-\mathrm{i}\delta_1\right)\left(\frac{\gamma}{2}-\mathrm{i}\delta_{ge}\right)+\lambda_1^2\right]
\left[\left(\frac{\kappa_2}{2}-\mathrm{i}\delta_2\right)\left(\frac{\gamma}{2}-\mathrm{i}\delta_{ge}\right)+\lambda_2^2\right]-\lambda_1^2\lambda_2^2}{
\left[\left(\frac{\kappa_1}{2}-\mathrm{i}\delta_1\right)\left(\frac{\gamma}{2}-\mathrm{i}\delta_{ge}\right)+\lambda_1^2\right]
\left[\left(\frac{\kappa_2}{2}-\mathrm{i}\delta_2\right)\left(\frac{\gamma}{2}-\mathrm{i}\delta_{ge}\right)+\lambda_2^2\right]-\lambda_1^2\lambda_2^2}
,   \nonumber     \\
&&t(\omega)=\frac{\lambda_1\lambda_2\sqrt{\kappa_1\kappa_2}\left(\frac{\gamma}{2}-\mathrm{i}\delta_{ge}\right)}{
\left[\left(\frac{\kappa_1}{2}-\mathrm{i}\delta_1\right)\left(\frac{\gamma}{2}-\mathrm{i}\delta_{ge}\right)+\lambda_1^2\right]
\left[\left(\frac{\kappa_2}{2}-\mathrm{i}\delta_2\right)\left(\frac{\gamma}{2}-\mathrm{i}\delta_{ge}\right)+\lambda_2^2\right]-\lambda_1^2\lambda_2^2}
\end{eqnarray}
\end{widetext}
are the reflection and transmission coefficients, respectively. If the frequency of the incident photon is large enough, $\{\delta_1, \delta_2, \delta_{ge}\}\gg\{\kappa_1, \kappa_2, \gamma, \lambda_1, \lambda_2\}$, the reflection coefficient $r(\omega)\simeq1$ and the transmission coefficient $t(\omega)\simeq0$.

If the TLS is decoupled from the two cavities, namely, $\lambda_1=0$ and $\lambda_2=0$, the cavities become independent as a result of Eqs.(\ref{eq1}). Therefore, the TLS behaviors just like a switch for controlling the input-output relation of the two cavities. The reflection coefficient is reduced to
\begin{eqnarray}
r_0(\omega)=\frac{-\frac{\kappa_1}{2}-\mathrm{i}\delta_1}{\frac{\kappa_1}{2}-\mathrm{i}\delta_1},
\end{eqnarray}
similar to input-output relation of the single-sided empty cavity mode, and the transmission coefficient reduces to $t_0(\omega)=0$. Consider a specific situation that $\kappa_1=\kappa_2=\kappa$, $\lambda_1=\lambda_2=\lambda$ and $\{\delta_1, \delta_2, \delta_{ge}\}\simeq0$, namely, the frequency of the incident photon is nearly resonated with the two cavities and the TLS is under the condition $\omega\simeq\omega_1\simeq\omega_2\simeq\omega_{ge}$. Therefore, the reflection and transmission coefficients become
\begin{eqnarray}
r_0=\frac{-\kappa\gamma}{\kappa\gamma+8\lambda^2},~~t_0=\frac{8\lambda^2}{\kappa\gamma+8\lambda^2},
\end{eqnarray}
respectively, satisfying $t_0=1+r_0$. We have $r_0=-1$ and $t_0=0$ if the TLS is decoupled from the two cavities, or $r_0\simeq0$ and $t_0\simeq1$ if the coupling strength satisfies $\lambda\gg\sqrt{\kappa\gamma/8}$.

\section{QUANTUM LOGIC GATES GENERATION ON PHOTONS}\label{S3}%(III)

\begin{figure}[!htbp]
\centering
\includegraphics[width=7.0cm,angle=0]{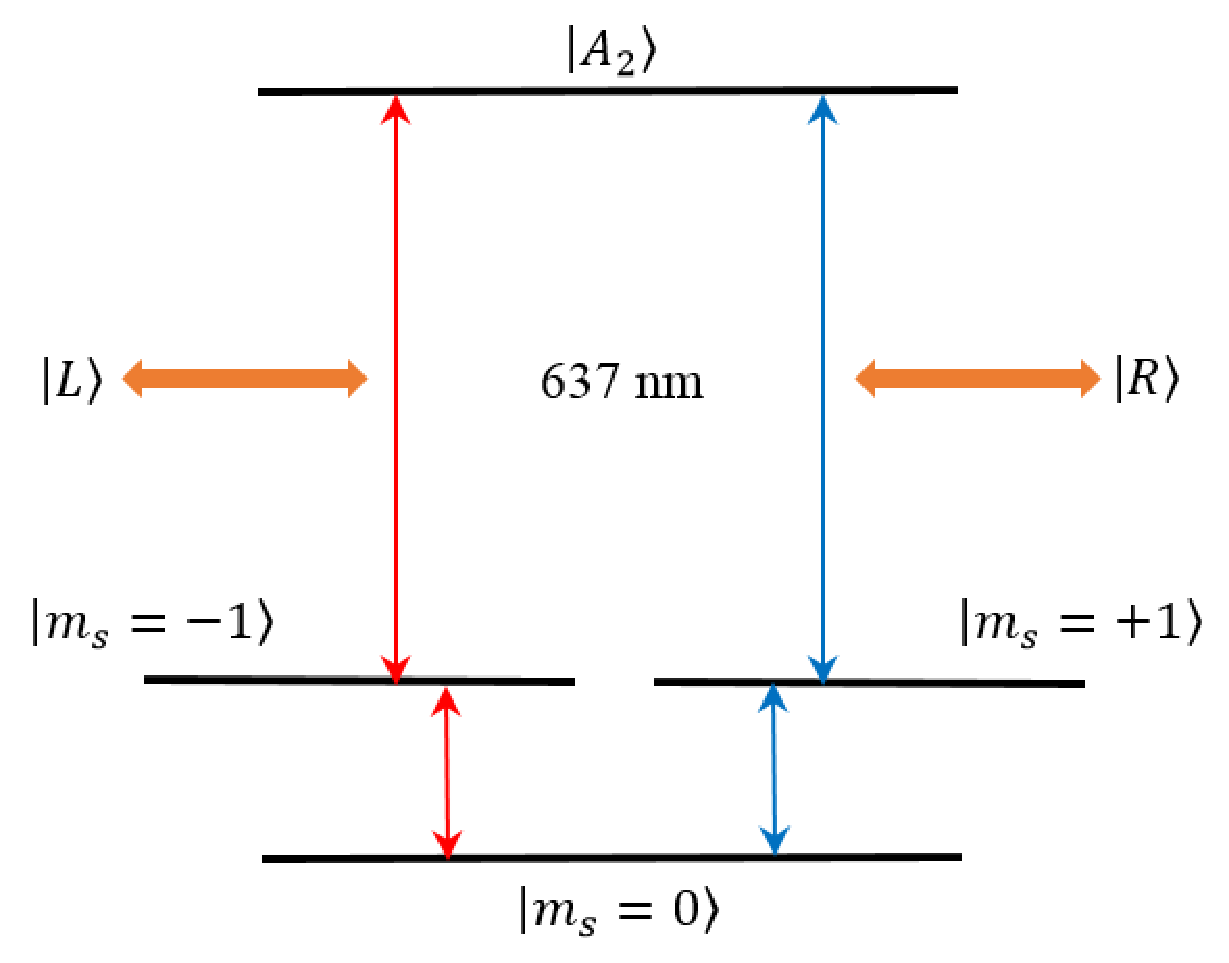}
\caption{(Color online) Schematic of energy levels of NV-center with an exited state $|A_2\rangle$ and the ground state. The ground state is a spin triplet with sublevels $|m_s=0\rangle$ and $|m_s=\pm1\rangle$. For convenience, $|m_s=\pm1\rangle\equiv|\pm\rangle$. The excited state $|A_2\rangle$ decays to $|-\rangle$ ($|+\rangle$) with left (right) circularly polarized photon $|L\rangle$ ($|R\rangle$).}\label{fig2}
\end{figure}

We next exploit the universal input-output relation for accomplishing relevant quantum computation tasks by using the NV-center instead of the TLS. As shown in Fig. \ref{fig2}, the ground state of single NV-center is the spin triplet $|m_s=0\rangle$ and $|m_s=\pm1\rangle=|\pm\rangle$. $|A_2\rangle=\frac{1}{\sqrt{2}}(|E_-\rangle|+\rangle+|E_+\rangle|-\rangle)$ is one of the excited states which are the eigenstates of the Hamiltonian of the NV-center, viewed as an auxiliary state, where $|E_{\pm}\rangle$ are the the orbital states. $|A_2\rangle$ has the stable symmetric properties. It is robust versus the limit of the small strain and magnetic fields. Additionally, the transition $|A_2\rangle\leftrightarrow|-\rangle$ absorbs and releases the left circularly polarized photon $|L\rangle$, while the transition $|A_2\rangle\leftrightarrow|+\rangle$ only concerns the right circularly polarized photon $|R\rangle$. These two transitions have the same probability. Generally, without external magnetic field, the left and right circularly polarized photons have the same wavelength 637 nm \cite{3NV1,3NV2}.

As the transitions $|A_2\rangle\leftrightarrow|-\rangle$ and $|A_2\rangle\leftrightarrow|+\rangle$ have the equal transition frequencies $\omega_-=\omega_+$ without the external magnetic field, it is assumed that the two transitions are resonated with the two cavities, namely, $\omega_-=\omega_1=\omega_2$ and $\omega_+=\omega_1=\omega_2$. We set also the coupling strengths between the two transitions, between the two cavities, and the spontaneous emission rates of the two transitions, to be the same, respectively. When the NV-center is in the state $|-\rangle$, the $|L\rangle$ photon incident from one port is interacted with the NV-center and transmitted to another port under the given condition $\lambda\gg\sqrt{\kappa\gamma/8}$. However, when the NV-center is in the state $|+\rangle$, the $|L\rangle$ photon will be reflected from the same port, since in this case it is equivalent to be decoupled from the two cavities simultaneously. The interactions between the $|L\rangle$ photon and the NV-center give rise to the following relations
\begin{eqnarray}
\left\{
\begin{aligned}
|L\rangle|-\rangle&\xrightarrow{\mathrm{NV}}|L\rangle|-\rangle \,\,\,(\mathrm{transmission}),  \\
|L\rangle|+\rangle&\xrightarrow{\mathrm{NV}}-|L\rangle|+\rangle \,\,\,(\mathrm{reflection}).  \\
\end{aligned}
\right.        \label{eq2}
\end{eqnarray}
For the incident $|R\rangle$ photon, if the NV-center is in the state $|-\rangle$, the $|R\rangle$ photon will be reflected from the same port. If the NV-center is in the state $|+\rangle$, the photon is interacted with the state and transmitted to another port under the same condition $\lambda\gg\sqrt{\kappa\gamma/8}$. Therefore, we have
\begin{eqnarray}
\left\{
\begin{aligned}
|R\rangle|-\rangle&\xrightarrow{\mathrm{NV}}-|R\rangle|-\rangle \,\,\,(\mathrm{reflection}),  \\
|R\rangle|+\rangle&\xrightarrow{\mathrm{NV}}|R\rangle|+\rangle \,\,\,(\mathrm{transmission}).  \\
\end{aligned}
\right.       \label{eq3}
\end{eqnarray}

\subsection{The generation of controlled-phase gate}\label{S3A}%(III-A)

\begin{figure}[!htbp]     %fig3
\centering
\includegraphics[width=8.0cm,angle=0]{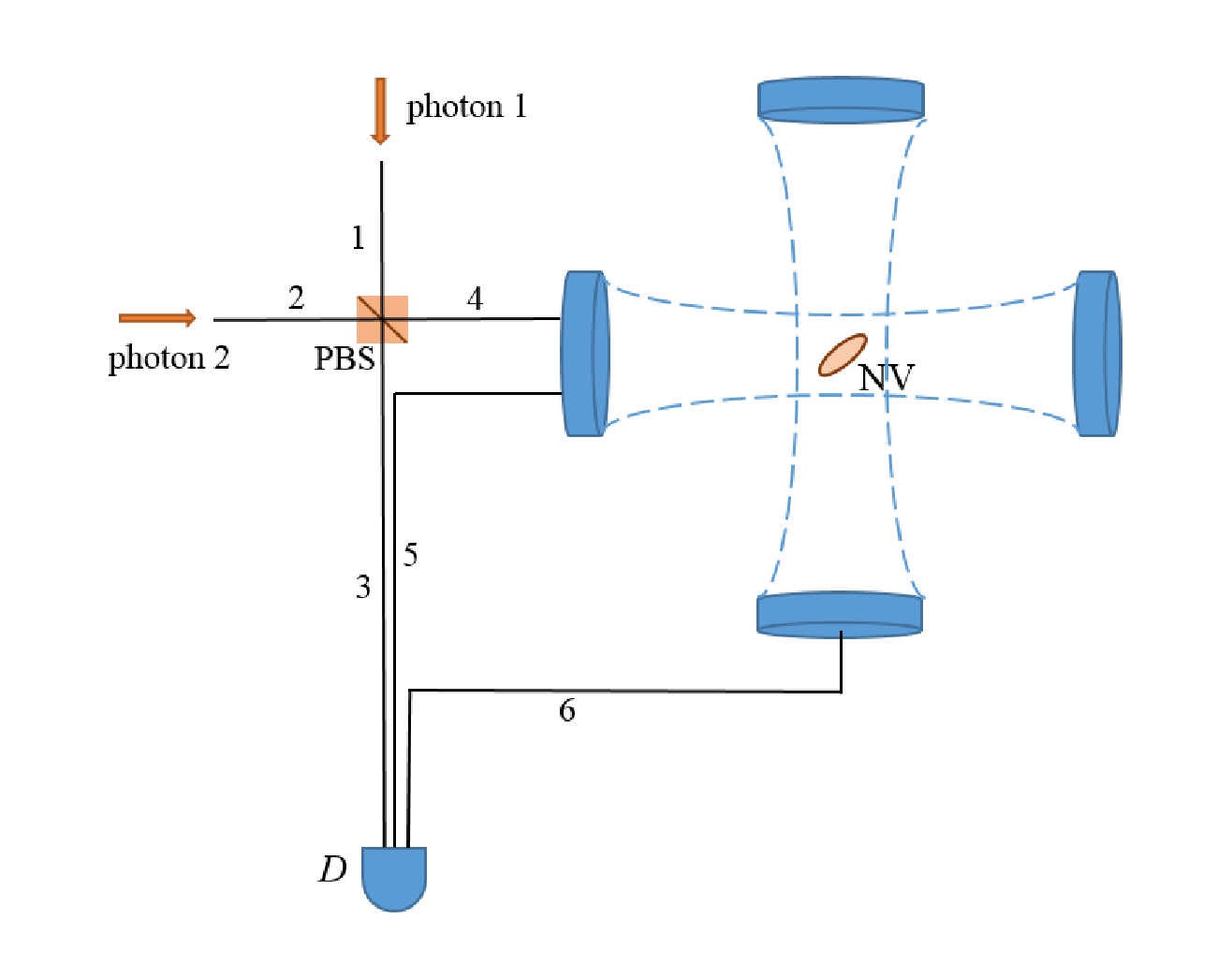}
\caption{(Color online) Schematic of the protocol of c-phase gate concerning two photons. The polarizing beam splitter (PBS) transmits the $|R\rangle$ circularly polarized photon and reflects the $|L\rangle$ circularly polarized photon, respectively. The photon 1 is incident from the line 1, then the photon 2 is incident from the line 2. The output state is detected by the detector $D$. The line 4 is only used for importing photons.}\label{fig3}
\end{figure}

The controlled-phase (c-phase) gate can be constructed concerning two incident photons. As shown in Fig. \ref{fig3}, the photons 1 and 2 incident from the line 1 and 2 are the control and target qubits, respectively. The state of the two photons are prepared as
\begin{eqnarray}
|\phi\rangle_1\!\!&=&\!\!a_1|R\rangle_1+b_1|L\rangle_1       \nonumber     \\
|\phi\rangle_2\!\!&=&\!\!a_2|R\rangle_2+b_2|L\rangle_2,
\end{eqnarray}
where $a_1$, $b_1$, $a_2$ and $b_2$ are complex parameters satisfying $|a_1|^2+|b_1|^2=1$ and $|a_2|^2+|b_2|^2=1$. As the auxiliary qubit, the NV-center is prepared in state
\begin{eqnarray}
|\mathrm{NV}\rangle=\frac{1}{\sqrt{2}}(|+\rangle+|-\rangle).
\end{eqnarray}
Therefore, the initial state of the two incident photons and the NV-center is
\begin{eqnarray}
|\psi\rangle_0=|\psi\rangle_{\mathrm{in}}\otimes|\mathrm{NV}\rangle,
\end{eqnarray}
where $|\psi\rangle_{\mathrm{in}}=|\phi\rangle_1\otimes|\phi\rangle_2$ is the input state of the two photons.

At the beginning, the photon 1 is incident from the line 1 shown in Fig. \ref{fig3}. Through the polarizing beam splitter (PBS), $|R\rangle_1$ is transmitted into the line 3 and detected by the detector $D$, namely, $|R\rangle_1$ is not interacted with the system. The $|L\rangle_1$ is reflected into the line 4 and interacted with the NV-center. The line 4 here is only used for importing photons. According to Eq.(\ref{eq2}), $|L\rangle_1|+\rangle$ can be reflected into the line 5 and $|L\rangle_1|-\rangle$ can be transmitted into the line 6. At last, both of them are detected by the detector $D$. The state becomes
\begin{eqnarray}
|\psi\rangle_1\!\!&=&\!\!a_1|R\rangle_1\frac{1}{\sqrt{2}}(|+\rangle+|-\rangle)|\phi\rangle_2     \nonumber   \\
&&\!\!+b_1|L\rangle_1\frac{1}{\sqrt{2}}(-|+\rangle+|-\rangle)|\phi\rangle_2.
\end{eqnarray}

Before the photon 2 is incident from the line 2, the Hadamard operation $H_{\mathrm{mad}}$,
\begin{eqnarray}
|+\rangle\!\!&\xrightarrow{H_{\mathrm{mad}}}&\!\!\frac{1}{\sqrt{2}}(|+\rangle+|-\rangle),  \nonumber   \\
|-\rangle\!\!&\xrightarrow{H_{\mathrm{mad}}}&\!\!\frac{1}{\sqrt{2}}(|+\rangle-|-\rangle),
\end{eqnarray}
is carried out on the NV-center. The state becomes
\begin{eqnarray}
|\psi\rangle_2=a_1|R\rangle_1|+\rangle|\phi\rangle_2-b_1|L\rangle_1|-\rangle|\phi\rangle_2.
\end{eqnarray}
Then the photon 2 is injected. Through the PBS, $|L\rangle_2$ is reflected into the line 3 and not interacted with the system; while $|R\rangle_2$ is transmitted into the line 4 and interacted with the NV-center. Based on Eq. (\ref{eq3}), $|R\rangle_2|-\rangle$ is reflected into the line 5 and $|R\rangle_2|+\rangle$ is transmitted into the line 6. So the state is given by
\begin{eqnarray}
|\psi\rangle_3\!\!&=&\!\!a_1|R\rangle_1(a_2|R\rangle_2+b_2|L\rangle_2)|+\rangle      \nonumber   \\
&&\!\!+b_1|L\rangle_1(a_2|R\rangle_2-b_2|L\rangle_2)|-\rangle.
\end{eqnarray}
At last, the Hadamard operation is performed again. The final state is
\begin{eqnarray}
|\psi\rangle_{\mathrm{f}}\!\!&=&\!\!\frac{|+\rangle}{\sqrt{2}}\bigg[a_1|R\rangle_1(a_2|R\rangle_2+b_2|L\rangle_2)     \nonumber    \\
&&\,\,\,\,\,\,\,\,\,\,\,+b_1|L\rangle_1(a_2|R\rangle_2-b_2|L\rangle_2)\bigg]     \nonumber    \\
&&\!\!+\frac{|-\rangle}{\sqrt{2}}\bigg[a_1|R\rangle_1(a_2|R\rangle_2+b_2|L\rangle_2)     \nonumber    \\
&&\,\,\,\,\,\,\,\,\,\,\,\,\,\,\,\,-b_1|L\rangle_1(a_2|R\rangle_2-b_2|L\rangle_2)\bigg].      \label{eq4}
\end{eqnarray}

If the output state of the NV-center detected by the detector $D$ is $|+\rangle$, no operation is performed on the two photons. If the output state of the NV-center is $|-\rangle$, the output state of the two photons is performed by the operation $\sigma_1^z=|R\rangle_1\langle R|-|L\rangle_1\langle L|$. In final, the output state of the two photons is
\begin{eqnarray}
|\psi\rangle_{\mathrm{out}}\!\!&=&\!\!a_1|R\rangle_1(a_2|R\rangle_2+b_2|L\rangle_2)   \nonumber   \\
&&\!\!+b_1|L\rangle_1(a_2|R\rangle_2-b_2|L\rangle_2).
\end{eqnarray}
In the process $|\psi\rangle_{\mathrm{in}}\rightarrow|\psi\rangle_{\mathrm{out}}$, the only phase shift is generated if and only if the states of the two photons are both $|L\rangle$, which is just the c-phase gate on two photons. The transformation matrix of the c-phase gate on two photons is
\begin{equation}
U_{\mathrm{c-phase}}=\left[\begin{array}{cccc}
 1 & 0 & 0 & 0   \\
 0 & 1 & 0 & 0   \\
 0 & 0 & 1 & 0   \\
 0 & 0 & 0 &-1
\end{array}\right]
\end{equation}
in the basis $\{|R\rangle_1|R\rangle_2, |R\rangle_1|L\rangle_2, |L\rangle_1|R\rangle_2, |L\rangle_1|L\rangle_2\}$.

\subsection{The generation of controlled-controlled-phase gate}\label{S3B}%(III-B)

\begin{figure*}[!htbp]    %fig4
\centering
\begin{tabular}{cc}
\includegraphics[width=6.0cm,angle=0]{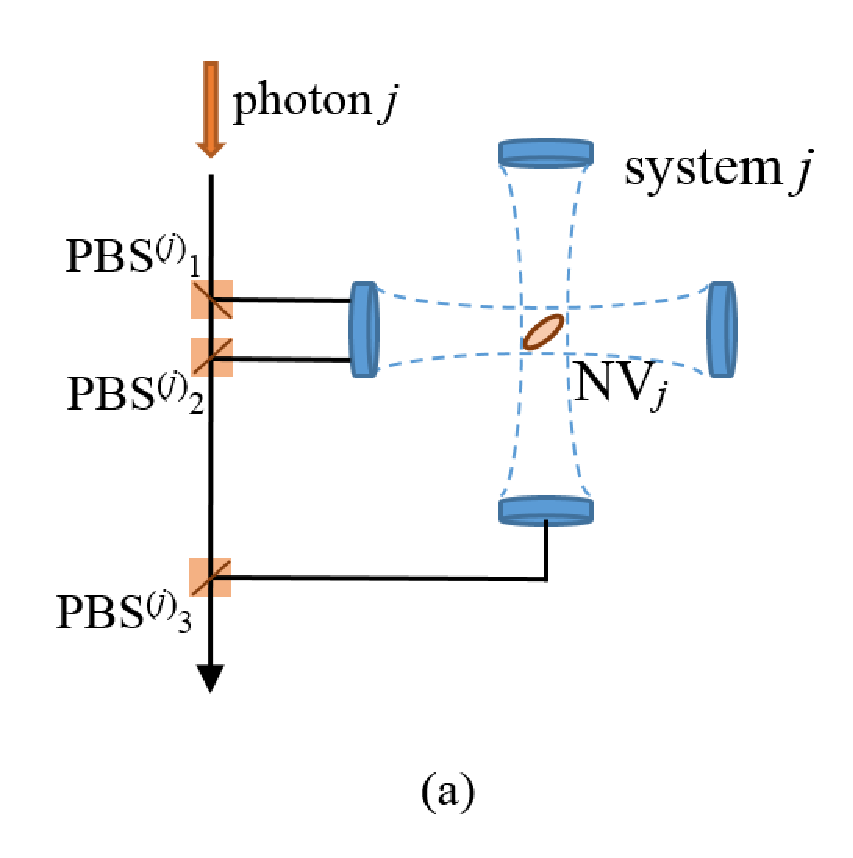} & \includegraphics[width=11.5cm,angle=0]{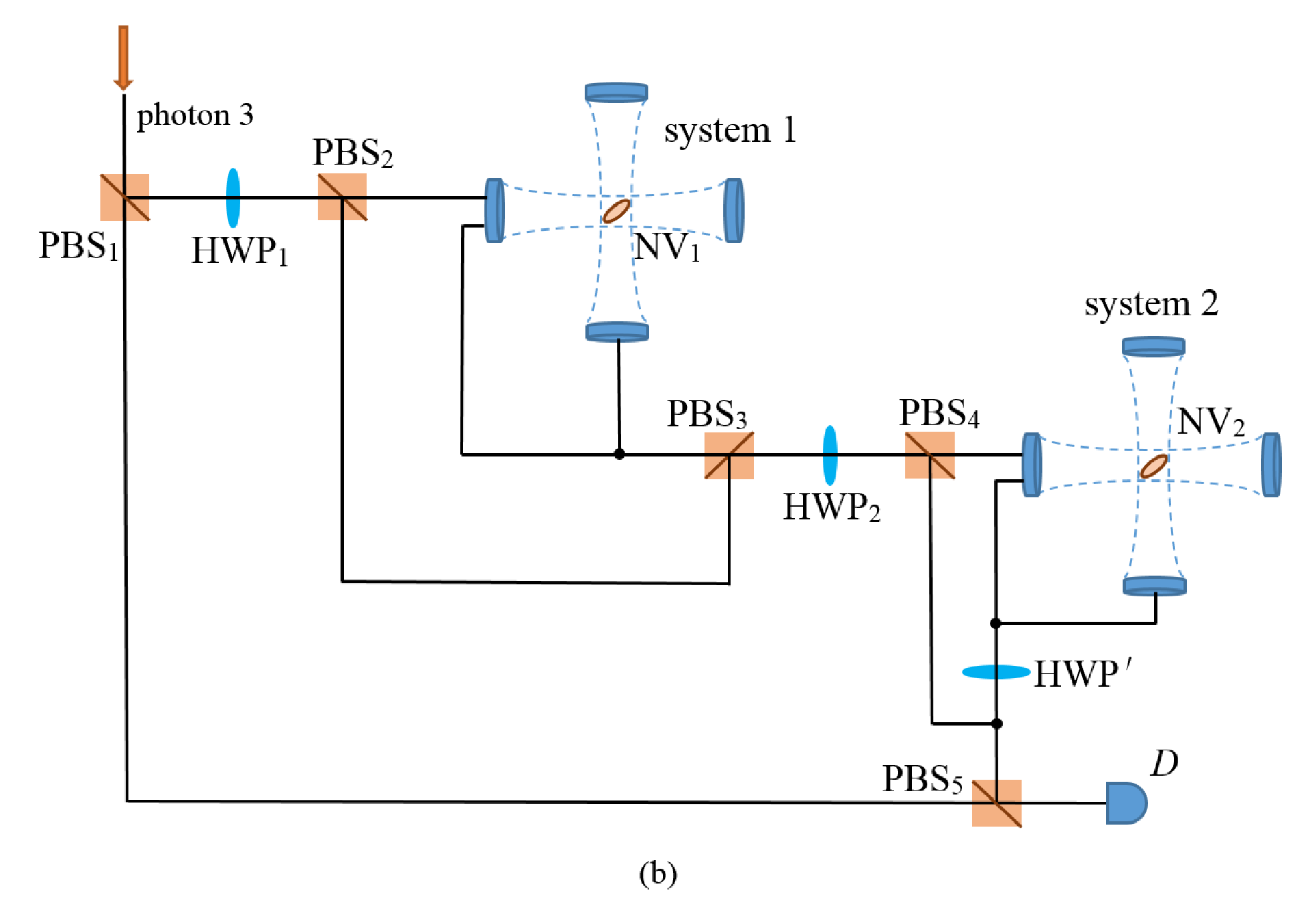}
\end{tabular}
\caption{In the schematic, the photon $j$ is firstly injected into the system $j$ ($j$=1,2) shown in the (a), prepared in the corresponding states of the photon $j$ and the system $j$. The photon 3 is incident from the input port (orange arrow) in the (b).}\label{fig4}
\end{figure*}

Now we consider how to generate the controlled-controlled-phase (cc-phase) gate on three photons, see Fig. \ref{fig4}. In this protocol, the photon 1 and photon 2 are both control qubits. The target qubit photon 3 is prepared in state
\begin{eqnarray}
|\phi\rangle_3=a_3|R\rangle_3+b_3|L\rangle_3
\end{eqnarray}
with $|a_3|^2+|b_3|^2=1$. Therefore, the initial state of three photons and two NV-centers is given by
\begin{eqnarray}
|\Psi\rangle_0=|\Psi\rangle_{\mathrm{in}}\otimes|\mathrm{NV}\rangle_1\otimes|\mathrm{NV}\rangle_2,
\end{eqnarray}
where $|\Psi\rangle_{\mathrm{in}}=|\phi\rangle_1\otimes|\phi\rangle_2\otimes|\phi\rangle_3$ is the input state of the three photons, two NV-centers NV$_1$ and NV$_2$ are auxiliary qubits prepared respectively in states
\begin{eqnarray}
|\mathrm{NV}\rangle_1\!\!\!&=&\!\!\!\frac{1}{\sqrt{2}}(|+\rangle_1+|-\rangle_1)      \nonumber   \\
|\mathrm{NV}\rangle_2\!\!\!&=&\!\!\!\frac{1}{\sqrt{2}}(|+\rangle_2+|-\rangle_2).
\end{eqnarray}

The states of the photon $j$ and the system $j$ ($j$=1,2) are prepared before the injection of photon 3. The photon 1 and the photon 2 are injected into the systems 1 and 2 at first, shown in Fig. \ref{fig4}(a), and the interaction results are detected by the detector $D$, respectively. Then the Hadamard operations are carried out on the two NV-centers. So the state becomes
\begin{eqnarray}
|\Psi\rangle_1\!\!\!&=&\!\!\!\left(a_1|R\rangle_1|+\rangle_1-b_1|L\rangle_1|-\rangle_1\right)     \nonumber    \\
&&\!\!\!\otimes\left(a_2|R\rangle_2|+\rangle_2-b_2|L\rangle_2|-\rangle_2\right)\otimes|\phi\rangle_3.
\end{eqnarray}

Then the photon 3 is incident from the input port, shown in Fig. \ref{fig4}(b). Through the PBS$_1$, $|R\rangle_3$ is transmitted and not interacted with the whole system, while $|L\rangle_3$ will be reflected and interacted with the rest components of the whole system. $|L\rangle_3$ passes the half-wave plate HWP$_1$ rotated by 22.5$^\circ$, which results in the two kinds of transformations $|R\rangle\rightarrow(|R\rangle+|L\rangle)/\sqrt{2}$ and $|L\rangle\rightarrow(|R\rangle-|L\rangle)/\sqrt{2}$. The state becomes
\begin{eqnarray}
|\Psi\rangle_2\!\!\!&=&\!\!\!\left(a_1|R\rangle_1|+\rangle_1-b_1|L\rangle_1|-\rangle_1\right)     \nonumber    \\
&&\!\!\!\otimes\left(a_2|R\rangle_2|+\rangle_2-b_2|L\rangle_2|-\rangle_2\right)             \nonumber     \\
&&\!\!\!\otimes\left[a_3|R\rangle_3+b_3\frac{1}{\sqrt{2}}\left(|R\rangle_3-|L\rangle_3\right)\right].
\end{eqnarray}

Next, the PBS$_2$ reflects the $|L\rangle_3$, and transmits the $|R\rangle_3$ that is interacted with the NV$_1$ in the system 1. By the PBS$_3$ and HWP$_2$, the state evolves to
\begin{eqnarray}
|\Psi\rangle_3\!\!\!&=&\!\!\!\big[\left(a_1|R\rangle_1|+\rangle_1-b_1|L\rangle_1|-\rangle_1\right) a_3|R\rangle_3    \nonumber   \\
&&+\left(a_1|R\rangle_1|+\rangle_1b_3|L\rangle_3+b_1|L\rangle_1|-\rangle_1b_3|R\rangle_3\right)\big]     \nonumber   \\
&&\!\!\!\otimes\left(a_2|R\rangle_2|+\rangle_2-b_2|L\rangle_2|-\rangle_2\right).
\end{eqnarray}

$|R\rangle_3$ is transmitted into the system 2 through the PBS$_4$, and interacted with the NV$_2$. The output state passes the HWP$^\prime$ rotated by -45$^\circ$, giving rise to the transformation $|R\rangle\leftrightarrow-|L\rangle$. $|L\rangle_3$ is interacted with the system 2 neither. Consequently, through the PBS$_5$ the final state of the three flying qubits and two NV-centers is
\begin{eqnarray}
|\Psi\rangle_{\mathrm{f}}\!\!\!&=&\!\!\!a_1|R\rangle_1|+\rangle_1\big[a_2|R\rangle_2|+\rangle_2\left(a_3|R\rangle_3+b_3|L\rangle_3\right)          \nonumber    \\
&&\,\,\,\,\,\,\,\,\,\,\,\,\,\,\,\,\,\,\,\,\,\,\,\,\,\,\,\,\,-b_2|L\rangle_2|-\rangle_2\left(a_3|R\rangle_3+b_3|L\rangle_3\right)\!\big]      \nonumber   \\
&&\!\!\!-b_1|L\rangle_1|-\rangle_1\big[a_2|R\rangle_2|+\rangle_2\left(a_3|R\rangle_3+b_3|L\rangle_3\right)          \nonumber    \\
&&\,\,\,\,\,\,\,\,\,\,\,\,\,\,\,\,\,\,\,\,\,\,\,\,\,\,\,\,\,-b_2|L\rangle_2|-\rangle_2\left(a_3|R\rangle_3-b_3|L\rangle_3\right)\!\big].  \label{eq5}
\end{eqnarray}

At last, through the detector $D$, the operations $\sigma_j^z=|R\rangle_j\langle R|-|L\rangle_j\langle L|$ ($j$=1,2) is applied on the two control qubits. Therefore, the final output state of two control qubits and one target qubit is
\begin{eqnarray}
|\Psi\rangle_{\mathrm{out}}\!\!\!&=&\!\!\!a_1|R\rangle_1\big[a_2|R\rangle_2\left(a_3|R\rangle_3+b_3|L\rangle_3\right)          \nonumber    \\
&&\,\,\,\,\,\,\,\,\,\,\,\,\,\,\,\,\,+b_2|L\rangle_2\left(a_3|R\rangle_3+b_3|L\rangle_3\right)\big]      \nonumber   \\
&&\!\!\!+b_1|L\rangle_1\big[a_2|R\rangle_2\left(a_3|R\rangle_3+b_3|L\rangle_3\right)          \nonumber    \\
&&\,\,\,\,\,\,\,\,\,\,\,\,\,\,\,\,\,+b_2|L\rangle_2\left(a_3|R\rangle_3-b_3|L\rangle_3\right)\big].
\end{eqnarray}

In the process $|\Psi\rangle_{\mathrm{in}}\rightarrow|\Psi\rangle_{\mathrm{out}}$, the phase shift is generated if and only if the states of the three photons are all in the left circularly polarized states $|L\rangle$, which is just the cc-phase gate on the three photons. The transformation matrix of the cc-phase gate on three photons is
\begin{equation}
U_{\mathrm{cc-phase}}=\left[\begin{array}{cc}
 I_7 & 0   \\
 0  & -1
\end{array}\right]
\end{equation}
in the basis $\{|R\rangle_1|R\rangle_2|R\rangle_3, |R\rangle_1|R\rangle_2|L\rangle_3, |R\rangle_1|L\rangle_2|R\rangle_3,\\ |R\rangle_1|L\rangle_2|L\rangle_3, |L\rangle_1|R\rangle_2|R\rangle_3, |L\rangle_1|R\rangle_2|L\rangle_3, |L\rangle_1|L\rangle_2|R\rangle_3,\\ |L\rangle_1|L\rangle_2|L\rangle_3\}$, where $I_7$ is the $7\times 7$ identity operator.

\section{THE FIDELITIES AND EFFICIENCIES OF THE PROTOCOLS}\label{S4}%(IV)

The ideal output states Eq.(\ref{eq4}) of the c-phase gate protocol and Eq.(\ref{eq5}) of the cc-phase gate protocol are attained by exploiting Eq.(\ref{eq2}) and Eq.(\ref{eq3}). Especially, the two transmission transformations are formulated under the ideal condition $\lambda\gg\sqrt{\kappa\gamma/8}$. Consider the realistic situation that the coupling strengths $\lambda$ between the NV-center and the two cavities are approximately  $\sqrt{\kappa\gamma/8}$. The two reflection transformations $|L\rangle|+\rangle\xrightarrow{\mathrm{NV}}-|L\rangle|+\rangle$ and $|R\rangle|-\rangle\xrightarrow{\mathrm{NV}}-|R\rangle|-\rangle$ are unchanged as a result of the similar decoupled regime. But the two transmission transformations become
\begin{eqnarray}
\left\{
\begin{aligned}
|L\rangle|-\rangle&\xrightarrow{\mathrm{NV}}\{t_0|L\rangle|-\rangle\}_{\mathrm{tr}}+\{r_0|L\rangle|-\rangle\}_{\mathrm{re}},   \\
|R\rangle|+\rangle&\xrightarrow{\mathrm{NV}}\{t_0|R\rangle|+\rangle\}_{\mathrm{tr}}+\{r_0|R\rangle|+\rangle\}_{\mathrm{re}},   \\
\end{aligned}
\right.
\end{eqnarray}
where the subscript $tr$ denotes the transmission interaction between the incident photon and the system, the subscript $re$ denotes that the incident photon can be reflected into the input port with certain probability. Therefore, the realistic output state in the c-phase gate generation shown in Fig. \ref{fig3} is given by
\begin{widetext}
\begin{eqnarray}
|\psi\rangle_{\mathrm{r}}\!\!\!&=&\!\!\!\frac{|+\rangle}{\sqrt{2}}\left[a_1|R\rangle_1\left(a_2p|R\rangle_2+b_2|L\rangle_2\right)      +b_1|L\rangle_1\left(a_2\frac{p^2+1}{2}|R\rangle_2-b_2|L\rangle_2\right)\right]   \nonumber  \\
&&\!\!\!+\frac{|-\rangle}{\sqrt{2}}\left[a_1|R\rangle_1\left(a_2p|R\rangle_2+b_2|L\rangle_2\right)
+b_1|L\rangle_1\left(a_2\frac{p^2-2p-1}{2}|R\rangle_2+b_2p|L\rangle_2\right)\right], \label{eq6}
\end{eqnarray}
where $p=t_0+r_0$. Similarly, the realistic output state in the cc-phase gate generation is
\begin{eqnarray}
|\Psi\rangle_{\mathrm{r}}\!\!\!&=&\!\!\!a_1|R\rangle_1\bigg[a_2|R\rangle_2\left(a_3|R\rangle_3|+\rangle_1|+\rangle_2
+\frac{\xi_3}{2}b_3|L\rangle_3|+\rangle_1|+\rangle_2\right)      \nonumber   \\
&&\,\,\,\,\,\,\,\,\,\,\,\,\,\,\,\,\,\,+b_2|L\rangle_2\left(a_3|R\rangle_3|+\rangle_1\frac{\xi_1|+\rangle_2-\xi_2|-\rangle_2}{2}
+b_3|L\rangle_3|+\rangle_1\frac{\xi_1\xi_3|+\rangle_2-2p\xi_2|-\rangle_2}{4}\right)\bigg]  \nonumber   \\
&&\!\!\!+b_1|L\rangle_1\bigg[a_2|R\rangle_2\left(a_3|R\rangle_3\frac{\xi_1|+\rangle_1-\xi_2|-\rangle_1}{2}|+\rangle_2
+b_3|L\rangle_3\frac{\xi_1\xi_3|+\rangle_1-2p\xi_2|-\rangle_1}{4}|+\rangle_2\right)   \nonumber   \\
&&\,\,\,\,\,\,\,\,\,\,\,\,\,\,\,\,\,\,\,\,\,+b_2|L\rangle_2\bigg(a_3|R\rangle_3
\frac{\xi_1|+\rangle_1-\xi_2|-\rangle_1}{2}\frac{\xi_1|+\rangle_2-\xi_2|-\rangle_2}{2}    \nonumber  \\
&&\,\,\,\,\,\,\,\,\,\,\,\,\,\,\,\,\,\,\,\,\,\,\,\,\,\,\,\,\,\,\,\,\,\,\,\,\,\,\,\,\,\,\,\,\,\,\,+b_3|L\rangle_3\frac{\xi_1^2\xi_3
|+\rangle_1|+\rangle_2-2p\xi_1\xi_2|+\rangle_1|-\rangle_2-2p\xi_1\xi_2|
-\rangle_1|+\rangle_2-2\xi_2^2|-\rangle_1|-\rangle_2}{8}\bigg)\bigg],\label{eq7}
\end{eqnarray}
\end{widetext}
where $\xi_1=p-1$, $\xi_2=p+1$ and $\xi_3=1+2p-p^2$.

The fidelity of the gates is generally defined by $F=|\langle\psi_{\mathrm{i}}|\psi_{\mathrm{r}}\rangle|^2$, where $|\psi_{\mathrm{i}}\rangle$ and $|\psi_{\mathrm{r}}\rangle$ are the ideal and realistic output states. In this work, the ideal output states $|\psi_{\mathrm{i}}\rangle$ are given by Eq.(\ref{eq4}) for the c-phase gate and Eq.(\ref{eq5}) for the cc-phase gate, while the realistic output states $|\psi_{\mathrm{r}}\rangle$ are given by Eq.(\ref{eq6}) and Eq.(\ref{eq7}), respectively. The average fidelities of the c-phase gate and cc-phase gate are given by
\begin{eqnarray}
\overline{F}_{\mathrm{cp}}\!\!\!&=&\!\!\!\frac{1}{(2\pi)^2}\int_0^{2\pi}\!\!\mathrm{d}\theta_1\int_0^{2\pi}\!\!\mathrm{d}\theta_2
|_{\mathrm{f}}\langle\psi|\psi\rangle_{\mathrm{r}}|^2,     \nonumber      \\
\overline{F}_{\mathrm{ccp}}\!\!\!&=&\!\!\!\frac{1}{(2\pi)^3}\int_0^{2\pi}\!\!\mathrm{d}\theta_1\int_0^{2\pi}\!\!\mathrm{d}\theta_2
\int_0^{2\pi}\!\!\mathrm{d}\theta_3|_{\mathrm{f}}\langle\Psi|\Psi\rangle_{\mathrm{r}}|^2.
\end{eqnarray}

The efficiency of a gate is defined as $\eta=\frac{n_{\mathrm{out}}}{n_{\mathrm{in}}}$, where $n_{\mathrm{in}}$ and $n_{\mathrm{out}}$ represent the input and output photon numbers, respectively. By taking into account all possible input and output states, the average efficiencies of the two gates are expressed as
\begin{eqnarray}
\overline{\eta}_{\mathrm{cp}}\!\!\!&=&\!\!\!\frac{1}{(2\pi)^2}\int_0^{2\pi}\!\!\mathrm{d}\theta_1\int_0^{2\pi}\!\!\mathrm{d}\theta_2
\frac{n_{\mathrm{out}}}{n_{\mathrm{in}}},     \nonumber   \\
\overline{\eta}_{\mathrm{ccp}}\!\!\!&=&\!\!\!\frac{1}{(2\pi)^3}\int_0^{2\pi}\!\!\mathrm{d}\theta_1\int_0^{2\pi}\!\!\mathrm{d}\theta_2
\int_0^{2\pi}\!\!\mathrm{d}\theta_3\frac{n_{\mathrm{out}}}{n_{\mathrm{in}}}.
\end{eqnarray}

Set $a_1=\cos\theta_1$, $b_1=\sin\theta_1$, $a_2=\cos\theta_2$, $b_2=\sin\theta_2$, $a_3=\cos\theta_3$ and $b_3=\sin\theta_3$. Therefore, by straightforward calculation, the fidelity and efficiency of the c-phase gate are given by
\begin{eqnarray}
\overline{F}_{\mathrm{cp}}\!\!&=&\!\!\frac{1}{64}\left(19p^2+26p+19\right)           \nonumber        \\
\overline{\eta}_{\mathrm{cp}}\!\!&=&\!\!\frac{1}{16}\big(p^4-2p^3+8p^2+2p+7\big).
\end{eqnarray}
Similarly, the fidelity and efficiency of the cc-phase gate are given by
\begin{eqnarray}
\overline{F}_{\mathrm{ccp}}\!\!&=&\!\!\frac{89p^4+244p^3+638p^2+708p+369}{2048}         \nonumber      \\
\overline{\eta}_{\mathrm{ccp}}\!\!&=&\!\!\frac{1}{512}(p^8-8p^7+40p^6-80p^5+154p^4       \nonumber     \\
&&\,\,\,\,\,\,\,\,\,\,\,-8p^3+144p^2+96p+173).
\end{eqnarray}

\begin{figure}
\centering
\includegraphics[width=10.0cm,angle=0]{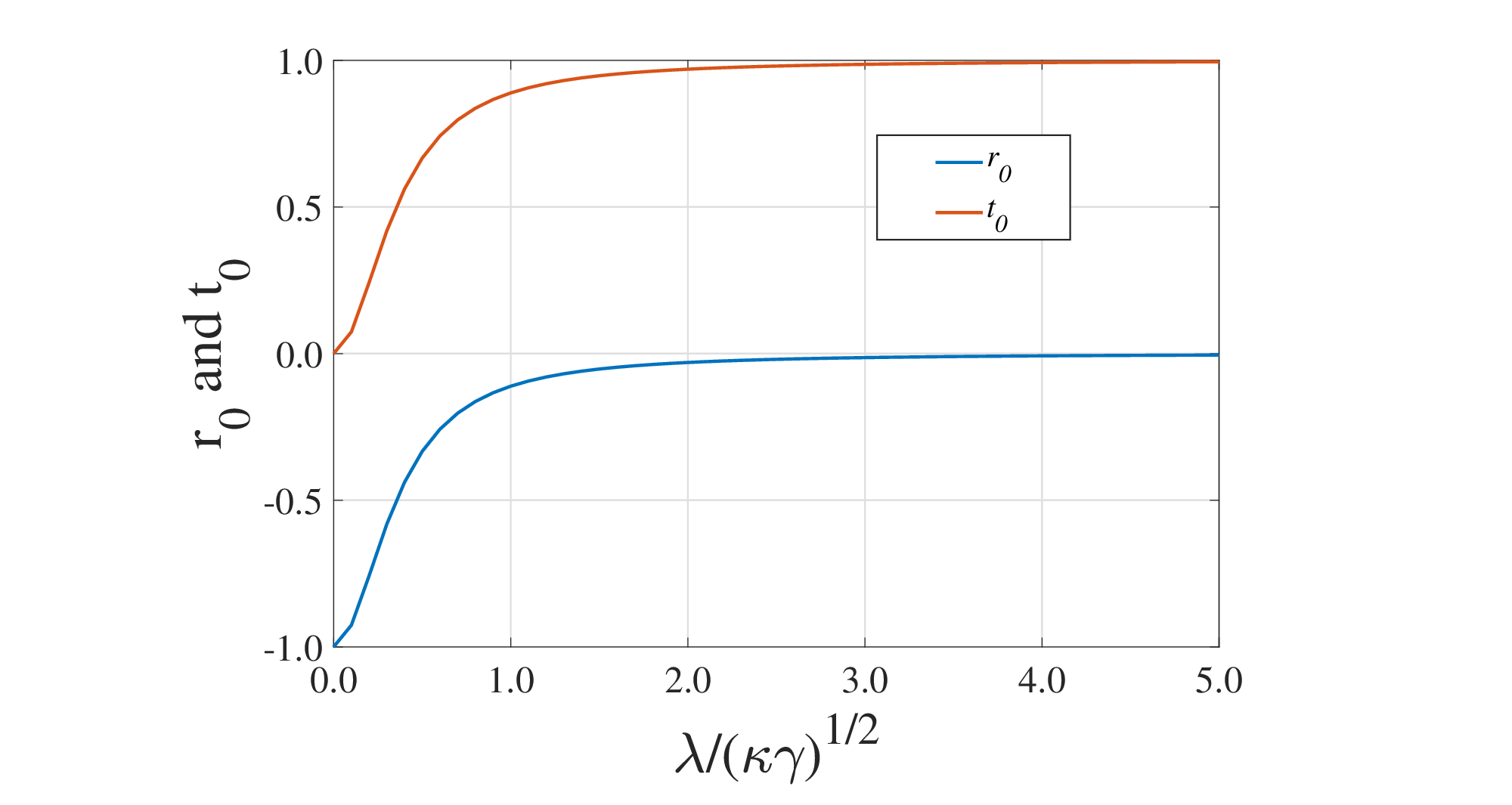}
\caption{(Color online) The value of the reflection coefficient $r_0$ and the transmission coefficient $t_0$ versus $\lambda/(\kappa\gamma)^{1/2}$.}\label{fig5}
\end{figure}

\begin{figure*}
\centering
\begin{tabular}{cc}
\includegraphics[width=9.0cm,angle=0]{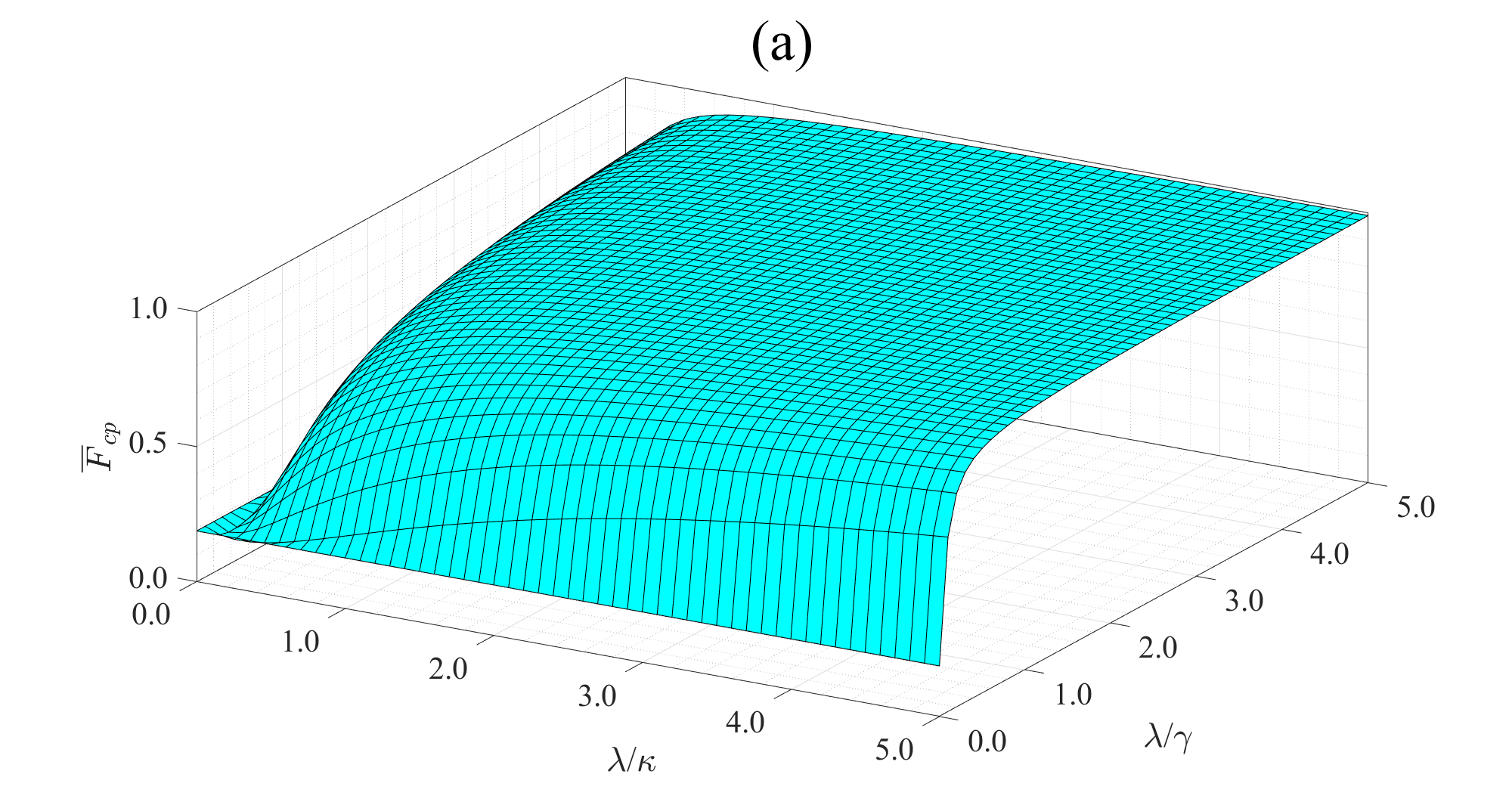} & \includegraphics[width=9.0cm,angle=0]{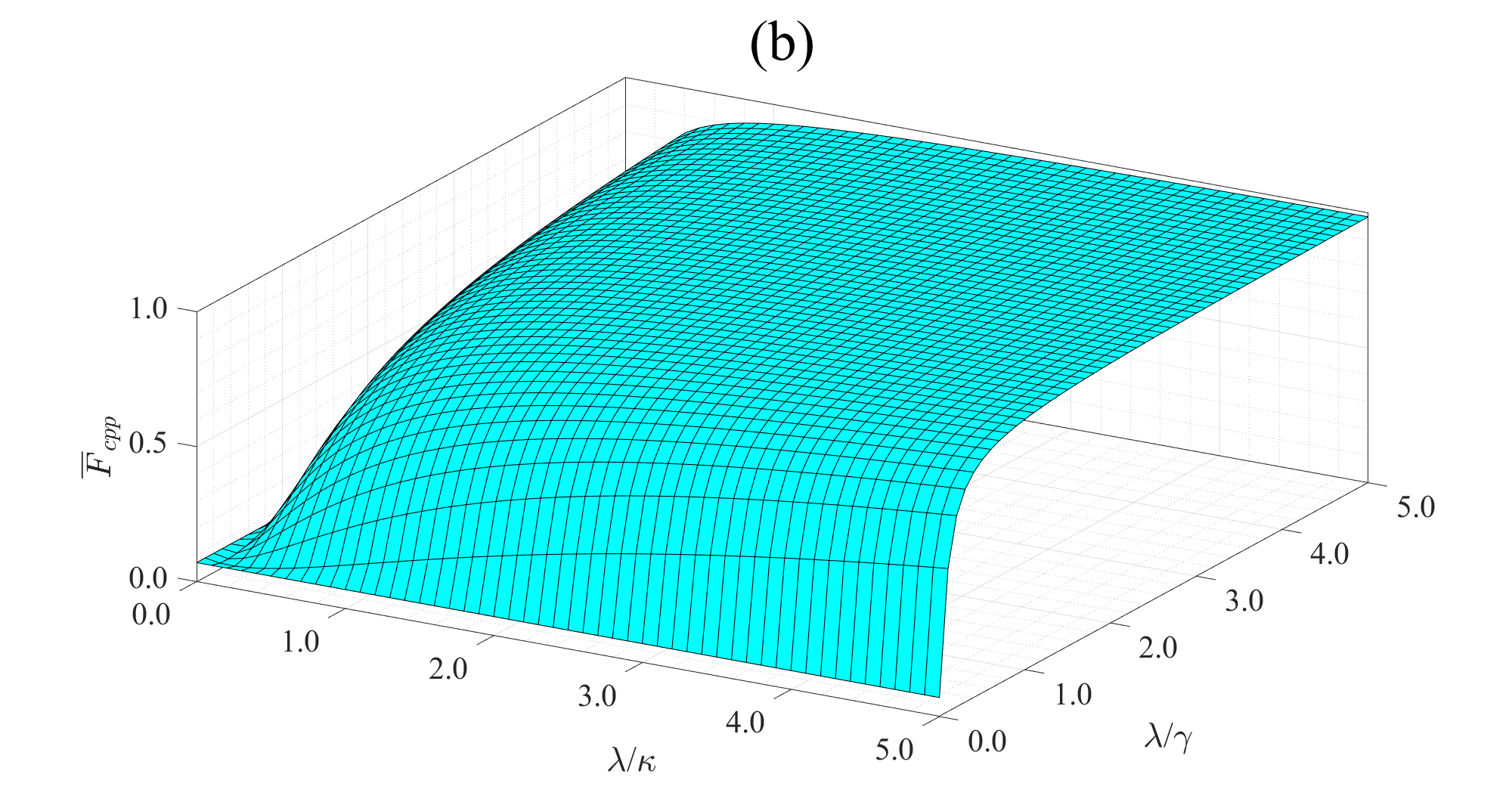}  \\
\includegraphics[width=9.0cm,angle=0]{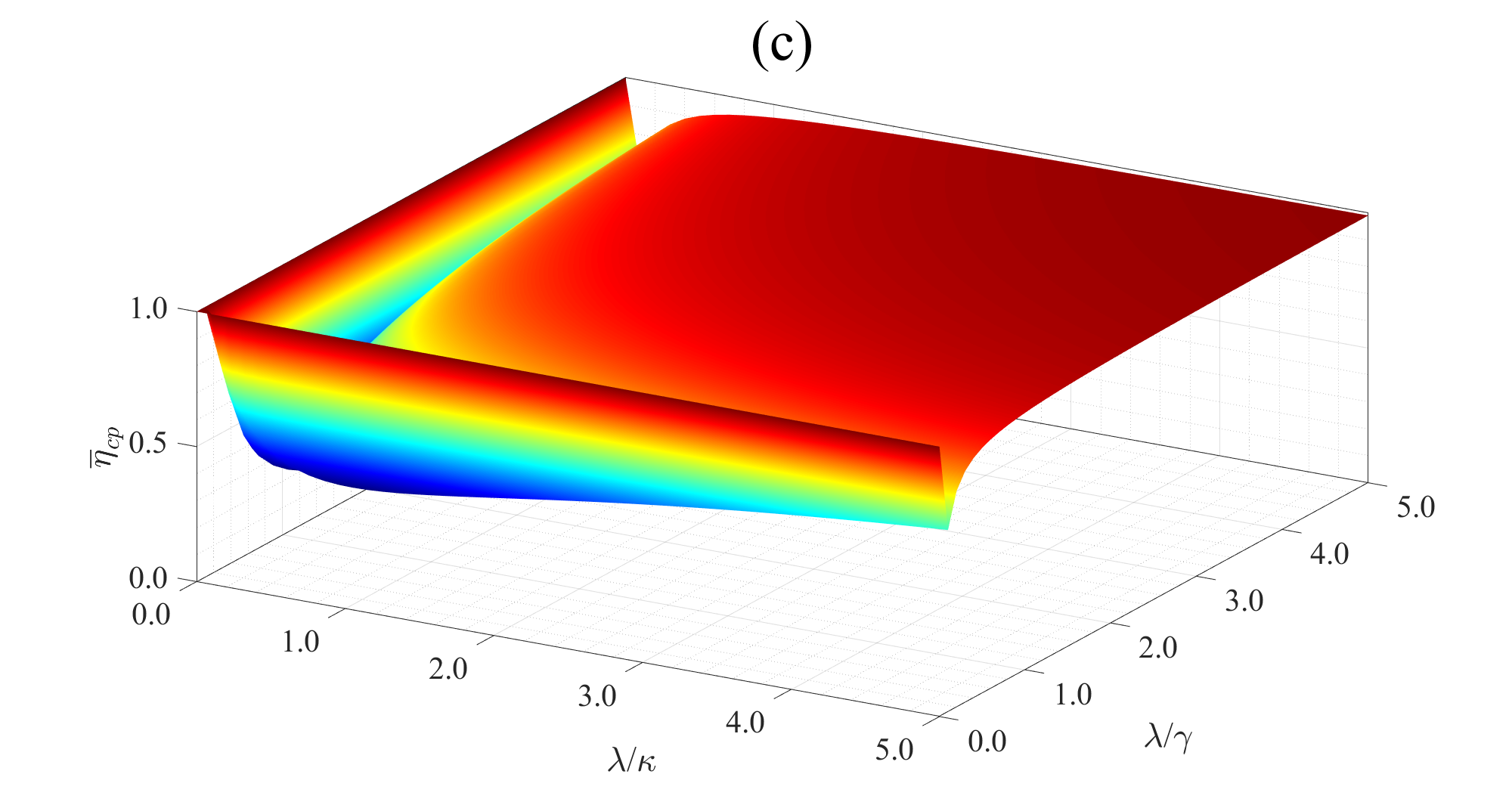} & \includegraphics[width=9.0cm,angle=0]{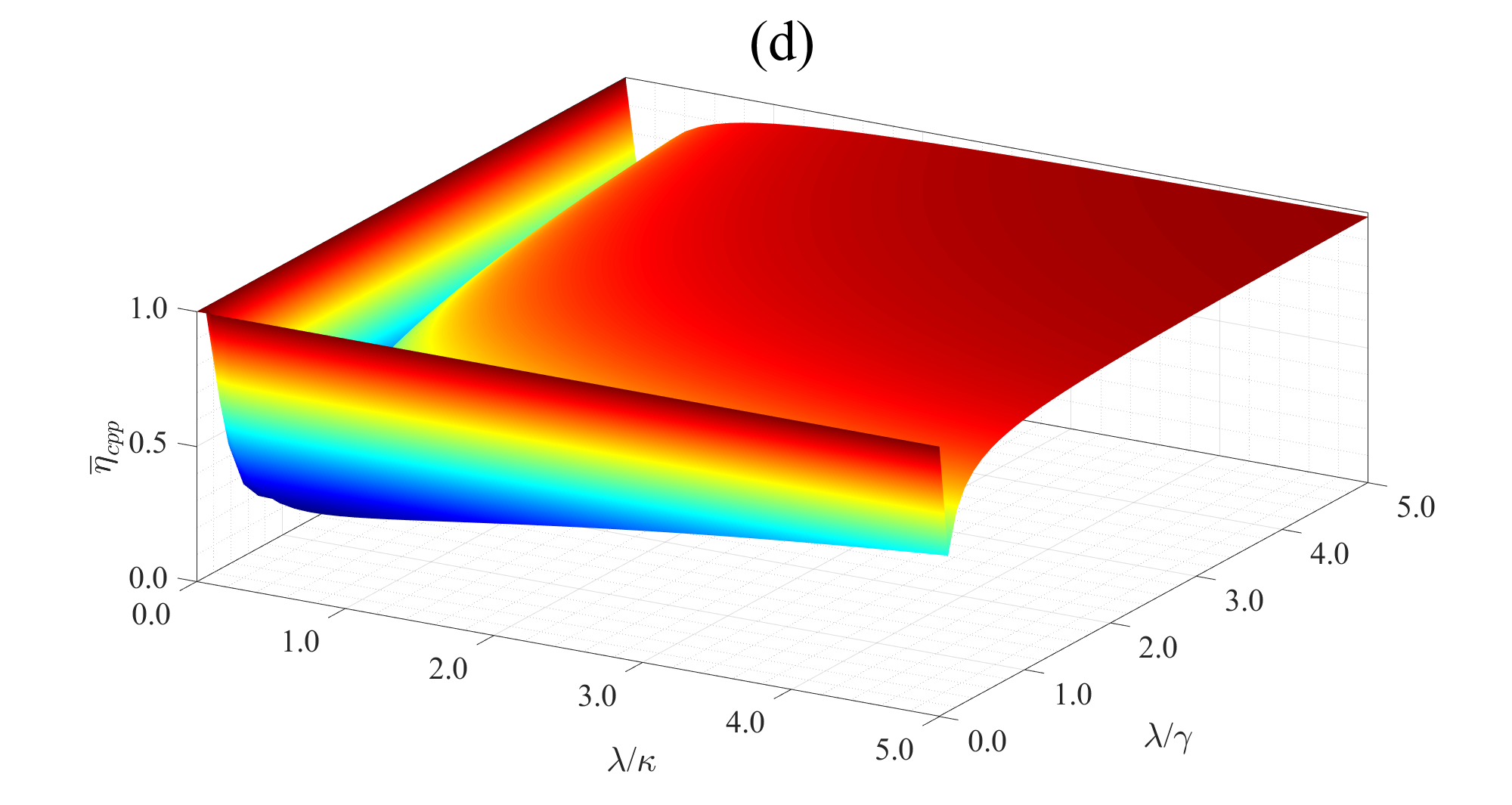}
\end{tabular}
\caption{ (Color online) The average fidelities and efficiencies of the two protocols versus $\lambda/\kappa$ and $\lambda/\gamma$: (a) The average fidelity of the c-phase gate $\overline{F}_{\mathrm{cp}}$; (b) The average fidelity of the cc-phase gate $\overline{F}_{\mathrm{ccp}}$; (c) The average efficiency of the c-phase gate $\overline{\eta}_{\mathrm{cp}}$; (d) The average efficiency of the cc-phase gate $\overline{\eta}_{\mathrm{ccp}}$. }\label{fig6}
\end{figure*}

The reflection coefficient $r_0$ and the transmission coefficient $t_0$ are attained at the specific condition: $\kappa_1=\kappa_2=\kappa$, $\lambda_1=\lambda_2=\lambda$ and $\omega\simeq\omega_1\simeq\omega_2\simeq\omega_{ge}$. They are functions of $\lambda/(\kappa\gamma)^{1/2}$, see Fig. \ref{fig5}. Given the decoherence of the NV-center and the decay of the cavity in our protocols, the average fidelities of the c-phase gate and cc-phase gate are detailed in Fig. \ref{fig6}(a) and (b), while the average efficiencies are shown in Fig. \ref{fig6}(c) and (d).

The average fidelities and efficiencies are functions of the parameters $\lambda$, $\kappa$ and $\gamma$. In recent years, much longer decoherence time of NV-center is attained in experiments. For example, Mizuochi \emph{et al.} \cite{4TNV1} observed that the decoherence time is up to 650 $\mu$s. Bar-Gill \emph{et al.} \cite{4TNV2} demonstrated that the decoherence time can approximately reach 0.6 s. Besides, the experiments \cite{4QC1,4QC2,4QC3,4QC4} showed that the optical cavities have much higher quality factors with 10$^{10}$-10$^{12}$, and the lifetime of the photon can be up to 2.5 ms at room temperature \cite{4QC4}. It should be noted that our two protocols can be accomplished in the weak coupling regime. Therefore, if $\lambda/\kappa=2$ and $\lambda/\gamma=2$, the average fidelities of the c-phase gate and cc-phase gate are 94.05$\%$ and 91.24$\%$, respectively. Meanwhile, the average efficiencies are 94.12$\%$ and 91.40$\%$. If $\lambda/\kappa=3$ and $\lambda/\gamma=3$ ($\lambda/\kappa=4$ and $\lambda/\gamma=4$), the fidelities are 97.28$\%$ and 95.96$\%$ (98.46$\%$ and 97.70$\%$), respectively.

Both of our protocols can also be accomplished in the strong coupling regime. For example, the decay rates of the optical cavities is chosen as $\kappa^{-1}$ = 20 $\mu$s \cite{4OCk}, and the decoherence time of NV-center is $\gamma^{-1}$ = 600 $\mu$s \cite{4TNV1}. Moreover, the coupling strengths are $\lambda/(2\pi)$ = 28 MHz \cite{4cN} between the NV-center and the optical cavities. Then the system is in the strong coupling regime as a result of $\lambda/\kappa\gg1$, $\lambda/\gamma\gg1$ and $\lambda\ll\{ \omega_-, \omega_+\}$, and the average fidelities of both protocols are $\overline{F}_{\mathrm{cp}}\simeq1$ and $\overline{F}_{\mathrm{cpp}}\simeq1$.

\section{CONCLUSION}\label{S5}%(V)

We have proposed a physical system consisting of two optical cavities coupled to a two-level system, which can be viewed as a double single-sided cavity system. The corresponding universal input-output relation, the reflection and transmission coefficients have been attained. As important applications, the controlled-phase gate and controlled-controlled-phase gate are constructed on photons with simple protocols by using the NV-center. Both protocols can be accomplished in the weak and strong coupling regimes. They give rise to much higher fidelities and efficiencies. Our approaches may highlight further investigations on physical realizations of quantum gates in quantum computation.

%\section*{Acknowledgements}

\end{document}